\documentclass{PoS}
\newcommand{\Psib}{\ensuremath{\overline{\Psi}}}

\title{Gauge-gravity duality -- Super Yang Mills Quantum Mechanics}

\ShortTitle{Gauge-gravity duality}

\author{\speaker{Simon Catterall}\\
        Department of Physics, Syracuse University, Syracuse NY 13244\\
        E-mail: \email{smc@physics.syr.edu}}

\author{Toby Wiseman\\
        Blackett Laboratory, Imperial College, London, SW7 2AZ\\
        E-mail: \email{t.wiseman@imperial.ac.uk}}

\abstract{We describe the conjectured holographic duality between Yang-Mills 
quantum mechanics and type IIa string theory. This duality allows us 
to use lattice Monte Carlo simulations to probe the physics of 
the gravitational theory - for example, at low energies it provides a 
computation of black hole entropy in terms of a sum over 
microstates of the dual gauge theory. Numerical results are presented of the 
4 supercharge theory at finite temperature}

\FullConference{The XXV International Symposium on Lattice Field Theory\\
		 July 30-4 August 2007\\
		 Regensburg, Germany}
\begin{document}

\section{Introduction}
The possible equivalence of string and gauge theory has long history
going back to birth of string theory as a possible theory of the
strong interactions and the recognition that Feynman diagrams of
perturbative large $N$ gauge theory could be naturally classified into a sum
over surfaces of given genus.

However, there has been renewed interest in the subject in recent
years stemming from Maldecena's discovery that the type II string theory
in anti-deSitter space could be described in terms of a particular
supersymmetric gauge theory in four dimensions \cite{adscft}. This theory has sixteen
real supersymmetries and gauge group $SU(N)$. This
so-called AdSCFT correspondence
has been tested most extensively in the limit $N\to\infty$ and
for low energies in which case the string theory reduces to classical
supergravity. However it is widely conjectured to be true away from these
limits. Perhaps surprisingly the gauge theory in question looks very different
from QCD -- it is a superconformal field theory with neither asymptotic freedom
nor confinement. 

Since then many other such {\it dualities} have been postulated corresponding
to varying the dimension of the gauge theory, adding additional matter fields
and breaking all or part of the supersymmetry
and conformal invariance. The dual gauge systems 
possess the common feature of residing on the boundary of the space on which
the string theory is defined
and are hence often referred to as giving a {\it holographic} representation of
the gravitational system. One such class of dualities arise naturally from
the original AdSCFT construction and consist of a mapping between
type II string theory containing $N$ $Dp$-branes and $(p+1)$ dimensional
supersymmetric gauge theories with gauge group $SU(N)$ \cite{dual}. In this talk we
will concentrate on perhaps the simplest of these systems
which describes the
dynamics of $D0$-branes in string theory in terms of super Yang Mills
quantum mechanics \cite{kabat}. 

We will be interested in this mapping at finite temperature corresponding
to a background string geometry in which the time coordinate is Euclidean
and periodic. At low energies the resultant
supergravity equations contain a black hole
with charge $N$ and metric 
\begin{equation}
ds^2=\alpha^\prime[-h(U)dt^2+h^{-1}(U)dU^2+
\frac{c^{\frac{1}{2}}\sqrt{\lambda}}{U^{\frac{3}{2}}}d\Omega^2_8\end{equation}
where ($\lambda=Ng_s{\alpha^\prime}^{-3/2}$) and the function $h(U)$ is given by
\begin{equation}
h(U)=\frac{U^{\frac{7}{2}}}{c^{\frac{1}{2}}\sqrt{\lambda}}
\left(1-\left(\frac{U_0}{U}\right)^7\right)\end{equation}
with $\alpha^\prime=l^2_{\rm string}$ is the (inverse) string tension
and $g_s$ the string coupling. Such a black hole has a
temperature and entropy given by 
\begin{equation}
\frac{T}{\lambda^{\frac{1}{3}}}\sim 
\left(\frac{U_0}{\lambda^{\frac{1}{3}}}\right)^{\frac{5}{2}}\qquad
S\sim N^2\left(\frac{U_0}{\lambda^{\frac{1}{3}}}\right)^{\frac{9}{2}}
\end{equation}
The holographic conjecture states that the
dual Yang-Mills model has $N$ colors
and is to be computed at the {\it same} temperature
with $\lambda$ identified as
usual 't Hooft coupling $\lambda=g_{YM}^2N$.
The statement of duality implies that the free
energies of both gauge theory and black hole are equal. Duality thus 
offers a way of understanding black hole entropy 
as arising from a counting of microstates in the dual gauge theory.

The analysis that leads to the duality
conjecture is only valid in which the string theory reduces to
supergravity.
This requires $Ng_s>>1$ and $\alpha^\prime\to 0$.
Thus the dual super Yang Mills quantum
mechanics is {\it strongly coupled}. This is rather a generic
feature of these gauge-gravity dualities -- typically the regime
in which the string theory is tractable corresponds to a strongly 
coupled large
$N$ gauge theory. Furthermore, in the quantum mechanics
case the dynamics of
the gauge theory depends only on the dimensionless coupling
$\beta=\frac{\lambda^{\frac{1}{3}}}{T}$ 
which then implies it is the low temperature behavior of
the gauge theory that provides a description of the dual black hole
thermodynamics.

It is not known what happens as we increase the temperature
of the gauge theory. Familiarity with
thermal gauge theories leads us to speculate that the
system can undergo a deconfining phase transition. Indeed, as we
shall show just such a transition occurs in the
quenched theory. In the gravitational language raising the
temperature
corresponds to increasing $\alpha^\prime$ and with it the
strength of classical string theory corrections. It is not
known for sure what happens in this case. It is possible that
the duality breaks down and no correspondence exists between the gauge
and gravity models. Or more exotic possibilities could exist with the
existence of a thermal phase transition in the gauge theory
signaling some sort
of phase transition in the gravitational system - for example a transition
from black holes to a hot gas of strings and branes. One of the goals of
our work will be to map out this gauge-gravity correspondence and to
understand the phase diagram of the string theory by conducting
numerical simulations of the gauge theory.

In this talk we will report on numerical simulations of a related model
with four supersymmetries. General arguments exist \cite{us} that this
model may lie in a similar universality class to its sixteen supercharge
cousin. This theory has also been studied recently in \cite{jun}.

\section{Action and supersymmetries}
The ${\cal Q}=16$ super Yang-Mills QM is
obtained by dimensional reduction of ${\cal N}=1$ SYM in
$D=10$ dimensions.
\begin{equation} S=\frac{N}{\lambda}\int^Rd\tau \left(\left(D_\tau X_i\right)^2-[X_i,X_j]^2+
2i\Psi^\alpha D_\tau\Psi_\alpha+
2\Psi^\alpha(\Gamma_i)^\beta_\alpha[X_i,\Psi_\beta]\right)\end{equation}
where $\Psi^\alpha$ is a 16-component spinor and $\Gamma_i=\gamma_0\gamma_i$
where $\gamma_i,i=0\ldots 10$ are the $D=10$ Dirac
matrices in Majorana representation. The bosonic
sector of the model consists of $9$ scalars $X_i$ and a gauge field $A$. All
fields take values in the adjoint representation of the gauge group.
The supersymmetries are given by
\begin{eqnarray}
\delta A&=&-2i\Psi_\alpha\epsilon^\alpha\\
\delta X_i&=&-2\epsilon^\alpha(\Gamma_i)^\beta_\alpha\Psi_\beta\\
\delta \Psi^\alpha&=&\frac{1}{2}\left(
(\Gamma_i)^\alpha_\beta iD_\tau X_i+\frac{1}{2}([\Gamma_i,\Gamma_j])^\alpha_\beta
[X_i,X_j]\right)\epsilon^\beta
\end{eqnarray}

After integration over the fermions one encounters a Pfaffian which on generic
scalar field backgrounds is complex. This renders simulation of the model
problematic. Because of this we have initially focused on a related
four supercharge model corresponding to the dimensional reduction of ${\cal N}=1$ super Yang
Mills in four dimensions. This model has received extensive attention
in the literature as it may be discretized on a lattice while preserving
part of the supersymmetry algebra exactly\footnote{A property it shares
with the ${\cal Q}=16$ model}\cite{latt2005}. The ${\cal Q}=4$ model has an action
similar to that shown above with the restrictions that there are now
just 3 scalars and the fermions are 4 component Majorana fields with Yukawa
couplings related to the corresponding four dimensional Dirac matrices
in Majorana representation.

\section{Lattice Theory}
The transition to a lattice theory is potentially delicate. Naive
discretizations generically break supersymmetry at the classical
level leading to the appearance of supersymmetry breaking counterterms
in the quantum effective action. Some of these may be relevant which leads
to a fine tuning problem as their associated couplings must then
be fine tuned to recover supersymmetry in the continuum limit.

However, super Yang-Mills quantum mechanics is, of course, 
super-renormalizable, which
implies that the continuum theory contains only a finite number of
superficially U.V divergent Feynman graphs and they
occur for small numbers of loops. It is only these terms that
can generate new supersymmetry violating interactions on the lattice.
This is the content of a powerful theorem on lattice Feynman diagrams
due to Reisz \cite{reisz} but it is easy to understand -- the lattice
propagators resemble continuum propagators for light modes and
only depart strongly for modes near the cut-off. In a theory with
no cut-off sensitivity these U.V modes are irrelevant and the lattice
diagram converges for vanishing lattice spacing to its continuum
counterpart. 

In our quantum mechanics case there is only one potentially
U.V sensitive diagram -- a one loop fermion tadpole correction to the
scalar field. However the gauge index structure of this term ensures
it must vanish even in the lattice theory. Thus we conclude
that any discretization of the continuum theory prohibiting doublers
will flow to the correct supersymmetric continuum limit
{\it without fine tuning}.

Hence we have employed
the following lattice action:
\begin{equation}S=\kappa\sum_\tau \left(
\sum_{i=1}^3\left(D^+ X_i\right)^2-\sum_{i<j}[X_i,X_j]^2+
\Psi^T{(1)} D^+\Psib^{(2)}+\Psi^T{(2)}D^-\Psi^{(1)}+
\Psi^T(\hat{\Gamma}_i[X_i,\Psi]\right)\end{equation}
Here $\Psi^{(1)},\Psi^{(2)}$ label the upper and lower components
of the fermion field.
The matrices $\hat{\Gamma}_i$ are essentially the $\Gamma_i$ up to an
redefinition of the Euclidean time direction in order that the
lattice kinetic term takes the explicitly antisymmetric form given
above \footnote{This trick happens automatically in the twisted formulations
described in \cite{latt2005}. In the ${\cal Q}=4$ case described here it
is possible to rewrite the Pfaffian
as the determinant of matrix operator of half the size and discretizations of
this operator can then choose a kinetic term proportional to the unit matrix. We
have also used this equivalent formulation in our numerical work} 

The lattice covariant derivative is just
\begin{equation}
D^+ f=U_1(t)f(t+\hat{t})U^\dagger_1(t)-f(t)\end{equation}
and the lattice coupling is given by
$\kappa=\frac{NM^3}{\beta^3}$ where we measure the inverse temperature
$\beta$ in units of the 't Hooft coupling $\lambda$ and $M$ is
the number of lattice
points. The continuum limit is simply $M\to\infty$ at fixed $\beta$. We 
also look for limiting behavior for large $N$.

\section{Simulation details}

It is not hard to show that
the Pfaffian arising after integration over the fermions
is real and positive semi-definite allowing it to be replaced by
the term 
$\left(M^\dagger M\right)^{\frac{1}{4}}$
where $M$ is the fermion operator described in the previous section
and we use antiperiodic thermal boundary conditions for the
fermions.
This latter weight is generated by an auxiliary integration over
pseudofermion fields $F,F^\dagger$ with action
\begin{equation}
S_{\rm PF}=F^\dagger \left(M^\dagger M\right)^{-\frac{1}{4}} F\end{equation}
Following the usual RHMC algorithm the latter is approximated
by a partial fraction expansion of the form
\begin{equation}
\frac{1}{x^{\frac{1}{4}}}\sim \alpha_0+\sum_{i=1}^Q
\frac{\alpha_i}{x+\beta_i}\end{equation}
where the $\alpha_i,\beta_i$ are computed using the remez algorithm
\cite{rhmc}. The resultant action is simulated using standard HMC techniques
together with a multimass CG-solver to solve the $Q$ linear systems
in a time independent of $Q$ \cite{susysims}.

We have simulated lattices with $5--12$ points at values of the
inverse temperature $\beta$ ranging from $0.01--3.0$. Typically
we have amassed between $10^3-10^4$ $\tau=1$ HMC trajectories
for systems with a number of colors in the range
$N=5--16$. 

Our primary observables are the absolute value of the trace of the
Polyakov loop $P$, its associated susceptibility
$\chi_P=\frac{1}{N}\left(<P^>-<P>^2\right)$ and the
mean energy of the system $E$. A simple scaling argument shows that
the latter is simply related to the expectation value of the
bosonic action $S_B$. We have also examined the mean extent of
the scalar fields given by 
$W=\frac{1}{\beta^{\frac{3}{4}}}\int d\mu |\mu| P(\mu)$. We have simulated
both the full supersymmetric theory and a quenched theory in which the
effects of dynamical fermions are removed.

\section{Results}
\begin{figure}
\begin{center}
\begin{tabular}{cc}
\includegraphics[height=40mm]{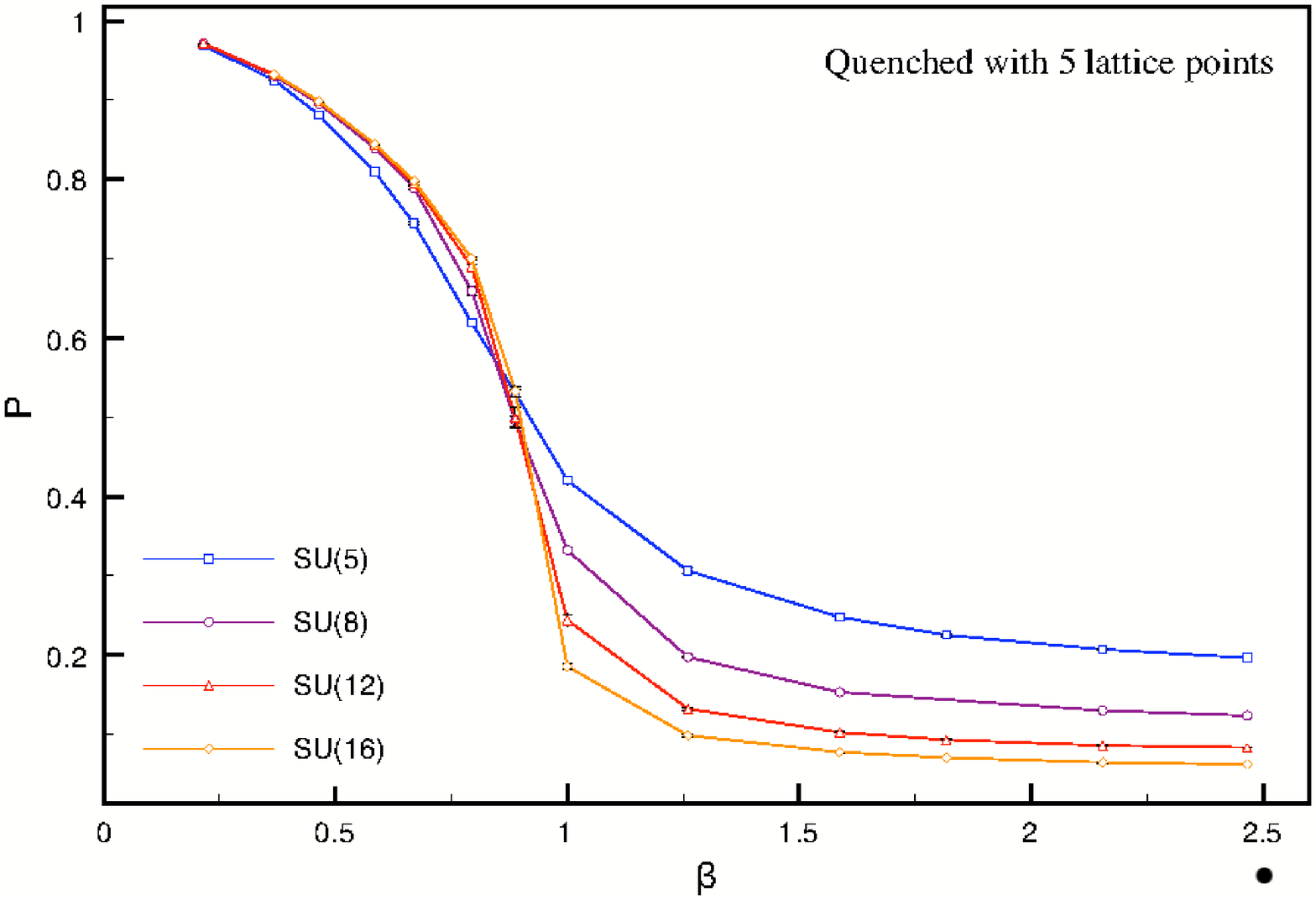} &
\includegraphics[height=40mm]{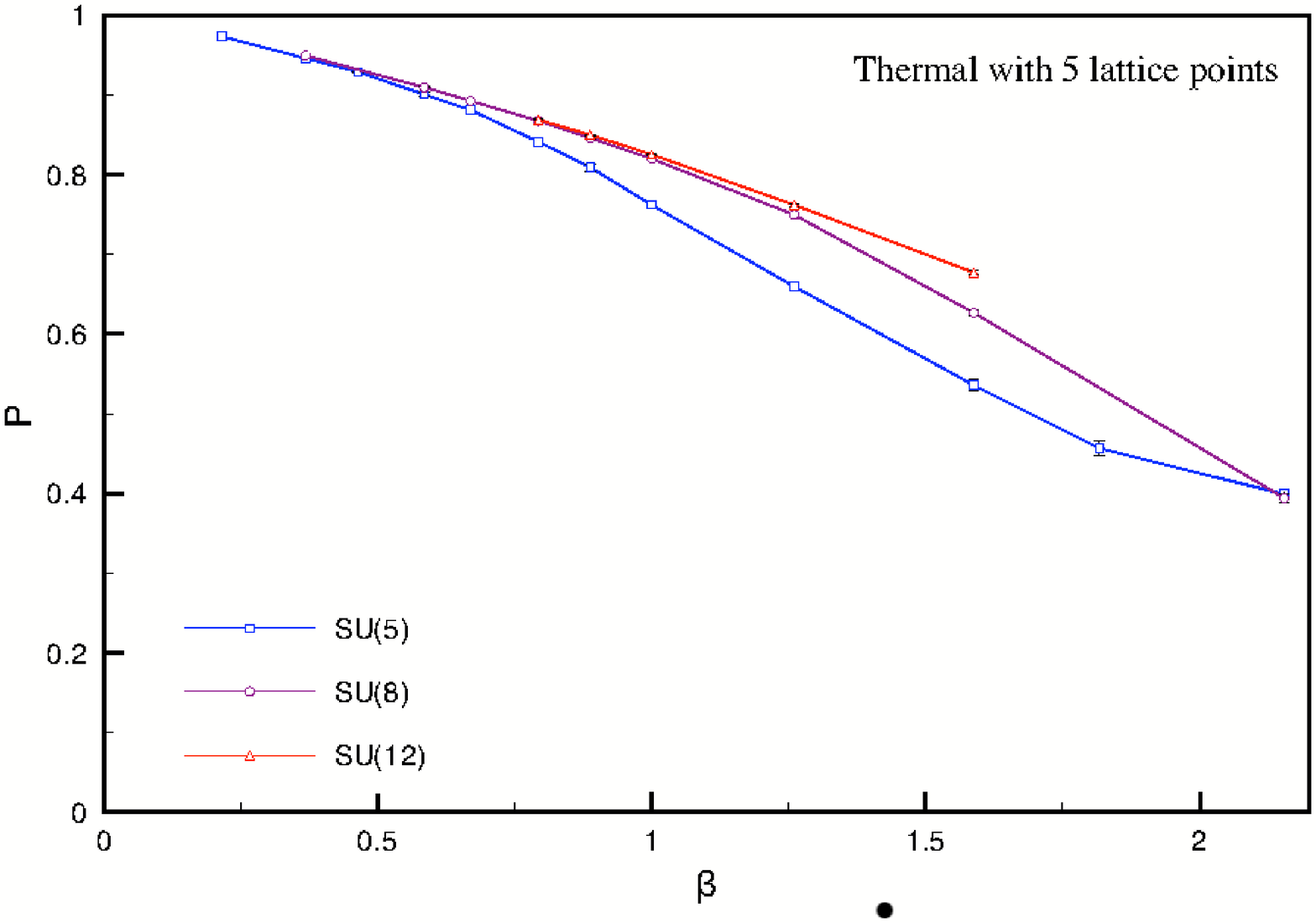}\\
Quenched  & Dynamical 
\end{tabular}
\end{center}
\caption{Polyakov line -- quenched and supersymmetric cases}
\label{poly}
\end{figure}

\begin{figure}
\begin{center}
\begin{tabular}{cc}
\includegraphics[height=40mm]{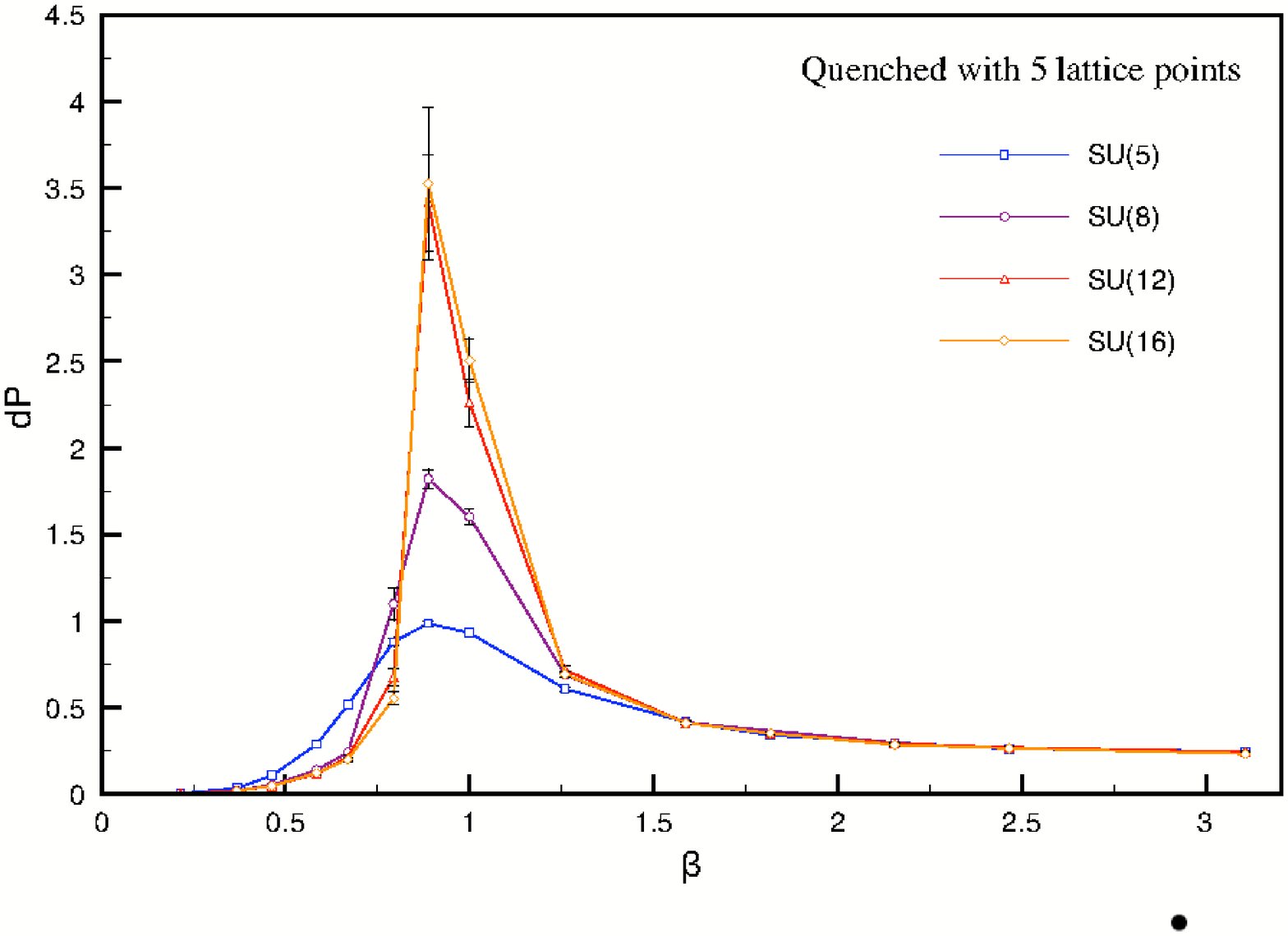} &
\includegraphics[height=40mm]{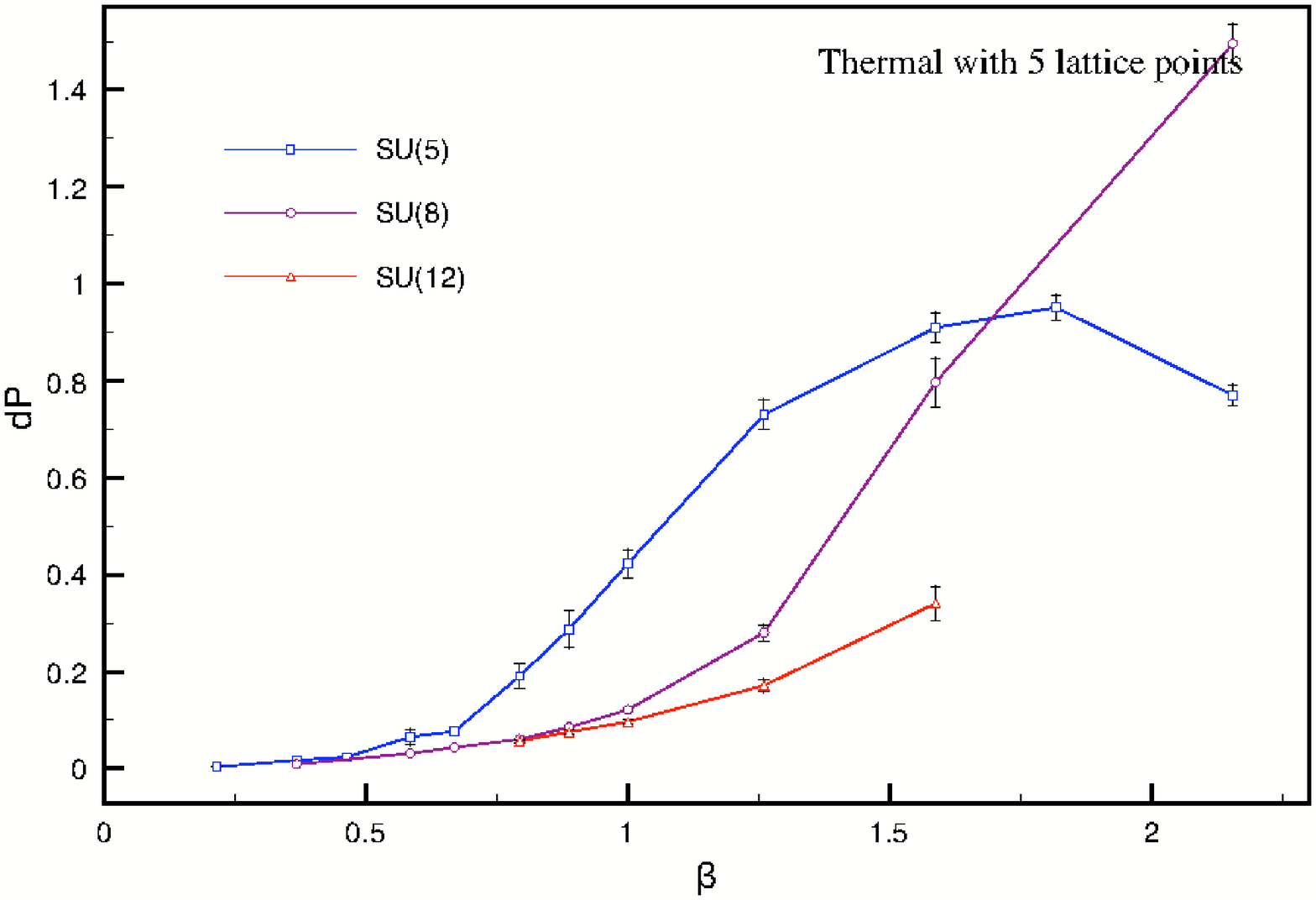}\\
Quenched  & Dynamical 
\end{tabular}
\end{center}
\caption{Polyakov susceptibility -- quenched and supersymmetric cases}
\label{sus}
\end{figure}

\begin{figure}
\begin{center}
\begin{tabular}{cc}
\includegraphics[height=40mm]{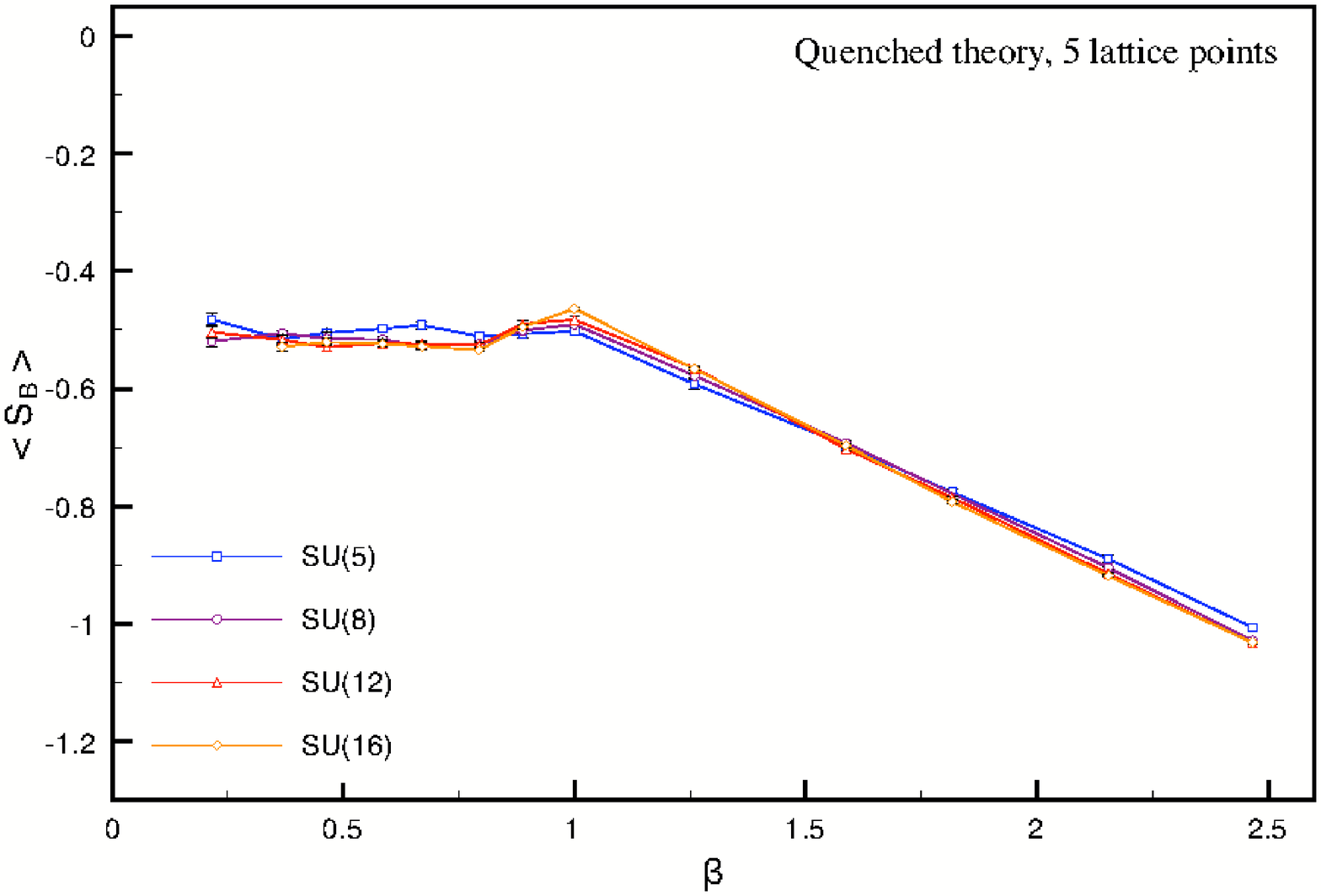} &
\includegraphics[height=40mm]{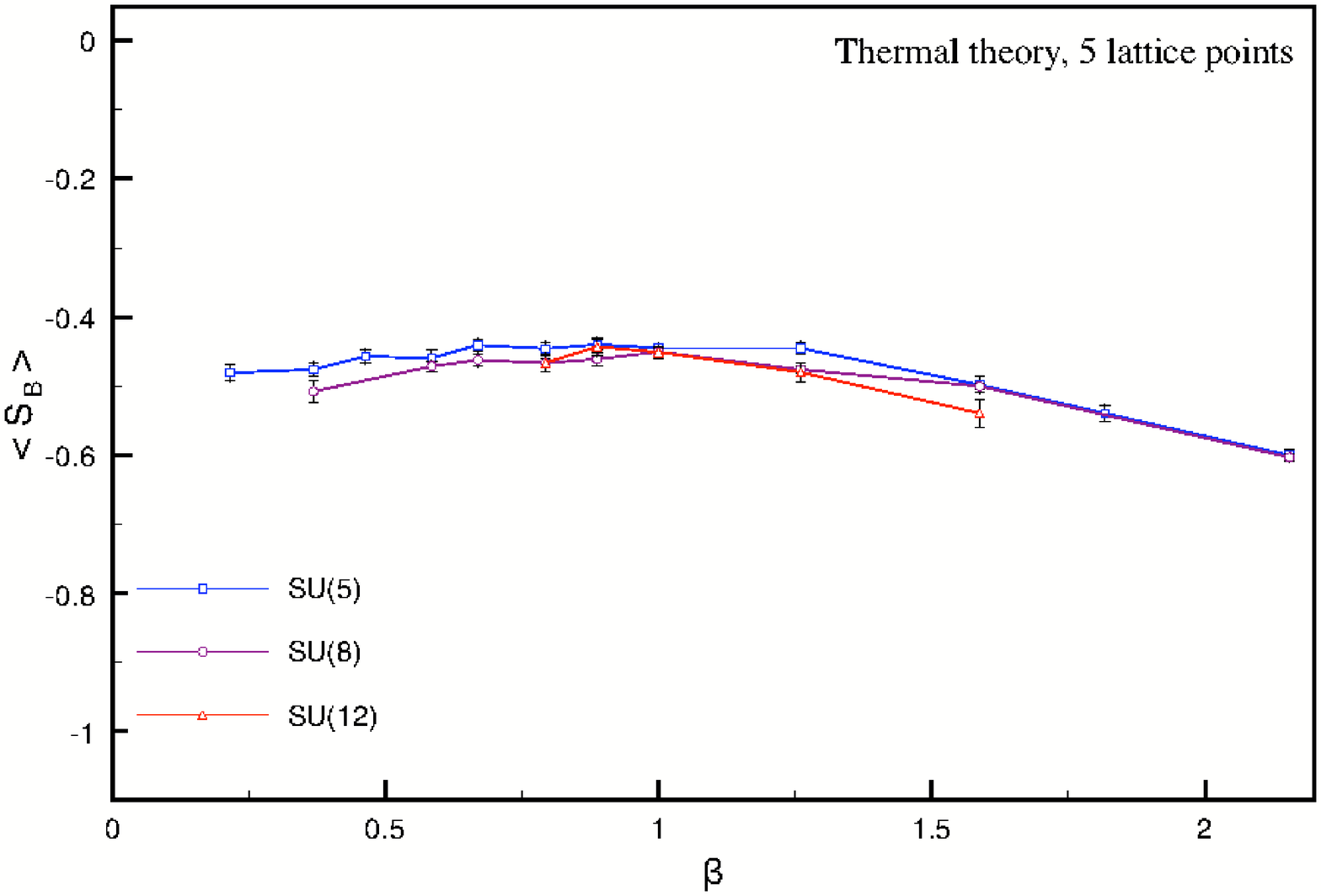}\\
Quenched  & Dynamical 
\end{tabular}
\end{center}
\caption{Mean energy -- quenched and supersymmetric cases}
\label{energy}
\end{figure}

Our results for the Polyakov line $P$ in both dynamical and quenched ensembles
are shown in figure~\ref{poly}
We see good evidence for 't Hooft scaling in both cases. In the quenched
theory we see evidence of a rapid crossover between a low temperature
confined phase and a high temperature deconfined phase. This crossover seems
to strengthen with increasing number of colors $N$. However no such crossover
is seen in the supersymmetric case and the system seems to exist in a 
single deconfined phase.
This conclusion is strengthened by looking at the susceptibility $\chi_P$
in figure~\ref{sus}
which shows a a sharp peak around $\beta_c\sim 0.9$
which grows with $N$ in the quenched theory but no such peak in the
supersymmetric theory - a very broad peak is seen at small $N$ and for
large $\beta$ but this seems to move rapidly to larger $\beta$ with
increasing $N$ and we conclude it will not survive the large $N$ limit.
Further evidence of this dramatic difference between quenched and supersymmetric
theories is seen in the energy $E$ shown in figure~\ref{energy}.
In the case of the quenched theory
the linear slope visible in $\beta<E>$ for large $\beta$ yields
the non-zero vacuum energy of the model  while the horizontal
regime below $\beta_c$ is consistent with classical equipartition
and the appearance of $O(N^2)$
weakly coupled degrees of freedom.
The dynamical system is quite
different -- we see good 't Hooft scaling but only weak $\beta$ dependence --
the large $\beta$ behavior is qualitatively similar to the quasi free
behavior seen at high temperature. Again the simplest interpretation
is that the deconfined phase persists for all finite temperatures. 
\section{Conclusions}
Holographic dualities offer the possibility of studying string
and quantum gravity theories using the tools of (supersymmetric) gauge theory. 
Typically the gauge theories that arise in these
contexts are strongly coupled which means that lattice methods are
useful.
Furthermore, low dimensional examples exist 
which allow for naive discretization
methods which nevertheless regain supersymmetry
with little or no fine tuning.
These models can be simulated relatively easily using modern
algorithms even in the
't Hooft large $N$ limit.

In this talk we have
presented results for a toy model with
just ${\cal Q}=4$ supercharges. Many features of this model
are thought to be common to the target ${\cal Q}=16$ model
but the toy model is easier to simulate primarily because it
possesses a real positive definite Pfaffian after integration
over the fermions. 

We have contrasted the supersymmetric simulations with
quenched simulations. In both cases good large $N$ scaling is seen.
In the quenched case the system exists in two phases in the limit
$N\to\infty$ -- a low temperature confined phase with non-zero
vacuum energy and a high temperature deconfined phase with $O(N^2)$
quasi free degrees of freedom.  In the supersymmetric case the
deconfined phase appears to extend to low temperatures. This is
in agreement with conjectures about the ${\cal Q}=16$ supercharge
theory \cite{us}. The latter theory has recently been studied using momentum
space methods in \cite{q=16} with interesting results. Perhaps most
encouraging is the observation that the phase fluctuations of the
Pfaffian were rather small over the temperature regime studied.
We are currently investigating the ${\cal Q}=16$
supercharge theory and hope to report
results soon \cite{us2}.

SC is supported in part by DOE grant
DE-FG02-85ER40237. TW is partly supported by a PPARC advanced
fellowship and a Halliday award. Simulations were performed using
USQCD resources at Fermilab.


\begin{thebibliography}{99}
\bibitem{adscft}J.Maldacena, Adv. Theor. Math. Phys. 2 (1998) 231.
\bibitem{dual} N. Itzhaki, Nissan, J. Maldacena, J. Sonnenschein and 
S. Yankielowicz, Phys. Rev. D58 (1998).
\bibitem{kabat} D. Kabat, G. Lifschytz and D. Lowe, Phys. Rev. D64, (2001)
124015.
\bibitem{us} Towards lattice simulation of the gauge 
theory duals to black holes and hot strings, S. Catterall and T. Wiseman,
arXiv:0706.3518 [hep-lat].
\bibitem{jun} Non lattice simulation for supersymmetric gauge theories in
one dimension, M. Hanada, J. Nishimura and S. Takeuchi,
arXiv:0706.1647 [hep-lat].
\bibitem{latt2005} K\"{a}hler-Dirac fermions and exact lattice supersymmetry,
S. Catterall, PoS LAT2005:006,2006. 
\bibitem{reisz} T. Reisz, Commun.Math.Phys.116:81,1988.
\bibitem{rhmc} M. Clark, PoS LAT2006:004,2006
\bibitem{susysims} S. Catterall,JHEP 0603:032,2006. 
\bibitem{q=16} K. Anagnostopoulos, M. Hanada, J. Nishimura and S. Takeuchi,
arXiv:0707.4454 [hep-th]
\bibitem{us2} S. Catterall and T. Wiseman, in preparation.

\end{thebibliography}
\end{document}